\documentclass{amsart}
\usepackage{amsmath}
\usepackage{amssymb}
\usepackage{amscd}
\usepackage[all]{xy}
\usepackage{xspace}

\def\S{{\mathfrak S}}

\def\cala{{\mathcal A}}

\def\hr{{ {H}_r(q)}} 
\def\PS{{\mathcal{PS}}}
\def\H{{\mathcal H}}
\def\I{{\mathcal I}}

\newtheorem{theorem}[subsection]{Theorem}
\newtheorem{proposition}[subsection]{Proposition}
\newtheorem{lemma}[subsection]{Lemma}
\newtheorem{corollary}[subsection]{Corollary}
\newtheorem{definition}[subsection]{Definition}
\newtheorem{remark}[subsection]{\bf Remark.}

\newcommand{\Vt}[1]{V^{\otimes #1}}
\def\glmn {\mathfrak{gl}_{m|n}}
\def\glm {\mathfrak{gl}_{m}}

\def \rvac {\!\! \left. \left. \right| \! 0 \right\rangle}

\newcommand{\beq}{\begin{equation}}
\newcommand{\eeq}{\end{equation}}
\newcommand{\ba}{\begin{array}}
\newcommand{\ea}{\end{array}}
\newcommand{\beqa}{\begin{eqnarray}}
\newcommand{\eeqa}{\end{eqnarray}}
\newcommand{\nn}{\nonumber \\}
\begin{document}

\author{Jean-Louis Loday \and Todor Popov}
\address{Institut de Recherche Math\'ematique Avanc\'ee\\
   CNRS et Universit\'e de Strasbourg\\
    7 rue R. Descartes\\
    67084 Strasbourg Cedex, France }
\email{loday@math.u-strasbg.fr}
\urladdr{www-irma.u-strasbg.fr/{$\sim$}loday/}

\address{Institute for Nuclear Research and Nuclear Energy, \\
          Bulgarian Academy of Sciences   \\
           bld. Tsarigradsko chauss\'ee 72\\
                BG-1784 Sofia,  Bulgaria}
\email{tpopov@inrne.bas.bg}

\title{Parastatistics Algebra, Young Tableaux  and  the Super Plactic Monoid}

\begin{abstract}
The parastatistics algebra  is a superalgebra with (even) parafermi
and (odd) parabose creation and annihilation  operators.
The states in the parastatistics Fock-like space are shown to be in one-to-one correspondence with the Super Semistandard Young Tableaux (SSYT) subject to further constraints. 
The deformation of the parastatistics algebra gives rise to a monoidal structure on the SSYT which is a super-counterpart of the plactic monoid.
\end{abstract}

\maketitle
\vspace{1cm}
\begin{flushright}
\begin{tabular}{l}
\`a Michel \\
d\'ecouvreur de senties
inconnus   \\ o\`u la beaut\'e math\'ematique
rejoint  \\ la simplicit\'e des lois de la physique.
\end{tabular}
\end{flushright}
\vspace{1cm}

\section{Introduction}	
Let $\mathcal A$ be an associative unital algebra. We denote by $\bullet$ and $[\ ,\ ]$ the associated symmetric and anti-symmetric operations: 
$$     x \bullet y = x y + y x \, , \qquad  \qquad             [ x , y ] = x y - y x  .$$
An easy lemma translates the associativity of $\mathcal A$ into the structure relations between
the induced products.

\begin{lemma}
\label{ll}
(M. Livernet, J.-L. Loday (unpublished))
The algebra $\mathcal A$ is associative
$$x(yz) = (xy)z $$
when the operation
 $[\, ,\, ] $ is a Lie bracket subject to the   relations
$$ \ba{rcl}
[ x \bullet y, z] &= &x \bullet [y , z ] +  [x , z] \bullet y \qquad \mbox{(Leibniz rule)}\, , \\[4pt]
[[x,y],z] &=& x \bullet (z\bullet y) - (x \bullet z ) \bullet y \, .
  \ea.
  $$
\end{lemma} 

\begin{definition} An algebra $\mathcal P$ is Poisson when the antisymmetric product $[ \, , \, ] $ and the symmetric product $\ast$  satisfy\\
(i) $[[x,y],z]+[[y,z],x]+ [[z,x],y]=0$ \quad \mbox{(Jacobi identity)} \\
(ii) $x \ast(y\ast z) = (x \ast y ) \ast z$, \\
(iii) $[ x \ast y, z] = x \ast [y , z ] + [x , z] \ast y \qquad \mbox{(Leibniz rule)} . $

\noindent
Here $[ \, , \, ] $ is the Poisson bracket and  $\ast$ stands for the commutative product of the algebra of the functions.
\end{definition}

A straightforward corollary of the Lemma (\ref{ll}) is the following one
\begin{corollary} Let the associative algebra $\cala$ be also a Poisson algebra
with commutative product $\ast$ coinciding with the symmetric product: 
$\bullet= \ast$. Then
the associativity of $\bullet$ implies  
\beq [[x,y],z] = 0 \, . 
\label{dc}\eeq
\end{corollary}


\section{Parastatistics algebra}
 
The relations of type (\ref{dc}) appear in the general quantization scheme due to  H.S. Green \cite{Green}. 
Instead of the canonical anticommutation relations 
between the creation $a^{\dagger}_{i}$ and annihilation $a_{j}$ modes of the fermions
\beq
[a_{i}, a^{\dagger}_{j}]_{+} = \delta_{ij} \ , \qquad 
[a_{i}, a_{j}]_{+}= 0 \ , \qquad  [a_{i}^{\dagger}, a_{j}^{\dagger}]_{+}=0 \ ,
\label{ccr}
\eeq
Green introduced a scheme coined parafermi quantization  based on the 
exchange relations 
\beqa 
\ba{rcccrcc} 
[[ a^{\dagger}_{i},a_{j} ], a^{\dagger}_{k}]&=& 2 \delta_{jk} 
a^{\dagger}_{i} &\quad
& [[ a^{\dagger}_{i},a_{j} ], a_{k}]&=& - 2 \delta_{i k} a_{j}
\\[4pt] [ [a^{\dagger}_{i}, a^{\dagger}_{j}],a^{\dagger}_{k} ] &=&0 & \quad &[
[a_{i}, a_{j}],a_{k} ] &=&0 \ea \label{PCR} 
\label{1}
\eeqa
 The associative algebra having relations (\ref{1}) will be referred to as creation-annihilation  parafermi algebra.
The bilinear canonical relations
imply the trilinear parafermi relations, (\ref{ccr})$\Rightarrow $(\ref{1})
thus the canonical quantization is a particular example of paraquantization.

The creation (and the annihilation) parafermi modes alone close a subalgebra with  double 
commutator relations as in (\ref{dc}).

More generally, 
for a system including both  odd and even degrees of freedom, 
the \textit{parastatistics relations} 
\beqa 
\ba{rcccrcc} 
[\![ [\![ a^{\dagger}_{i},a_{j} ]\!], a^{\dagger}_{k}]\!] &=& 2 \delta_{jk} 
a^{\dagger}_{i} &\quad
& [\![ a_{k},[\![ a^{\dagger}_{j},a_{i} ]\!]]\!] &=&  2 \delta_{j k} a_{i}
\\[4pt] [\![[\![a^{\dagger}_{i}, a^{\dagger}_{j}]\!],a^{\dagger}_{k} ]\!] &=&0 & \quad &
[\![ a_{k},[\![a_{j}, a_{i}]\!] ]\!] &=&0 \ea 
\label{super}
\eeqa
define the  superalgebra 
where $[\![x,y]\!]:= xy - (-1)^{\hat{x}\hat{y}} yx$ is a  Lie superbracket,
the parabose operators are odd, and the parafermi ones are 
even generators (note that here the grading is  the opposite to the usual one in which bose are even and fermi are odd).
 The Lie superalgebra $\mathcal L$ closed from the creation parastatistics modes $a_i^{\dagger}$  is $2$-nilpotent in view of the  relation 
$
[\![[\![a^{\dagger}_{i}, a^{\dagger}_{j}]\!],a^{\dagger}_{k} ]\!] =0 
$, cf. (\ref{super}),
thus for the  Lie superalgebra $\mathcal L$ we have
$$
\mathcal L = V \oplus [\![ V, V ] \!].
$$
\begin{definition}
Let us  denote by $V$ the  vector superspace of dimension $m|n$
spanned by the  even ($\hat{i}=0$)  parafermionic
   and  odd ($\hat{i}=1$) parabosonic  creation   operators
$V=V_0 \oplus V_1
\cong \mathbb C^{m|n}$ and we
suppose  $V_0=\bigoplus_{i=1}^m \mathbb C a_i^{\dagger}$
and $V_1=\bigoplus_{i=m+1}^{m+n} \mathbb C a_i^{\dagger}:=
\bigoplus_{i=\bar{1}}^{\bar{n}} \mathbb C a_{i}^{\dagger}$.

The {\it creation parastatistics algebra} $PS(V)$ is the universal enveloping algebra of the 
Lie superalgebra $\mathcal L$ ,
\beq
PS(V) =U(\mathcal L)= T(V)/I(V)  \qquad I(V)=([\![V,[\![V,V]\!]_{\otimes} ]\!]_{\otimes})
\eeq
that is, the factor of
the tensor algebra $T(V)$ by the  ideal $I(V)$ generated by the double supercommutators \cite{D-VP}.
\end{definition}
From the Poincar\'e-Birkhoff-Witt theorem for Lie superalgebras  one gets \cite{D-VP}
\beq
\label{PBW}
PS(V)= U(\mathcal L)\cong \mathbb S(V)\otimes \mathbb S([\![ V, V ] \!])
\eeq
where $\mathbb S(A)$ is the symmetric superalgebra generated from $A$ (see below).

\subsection{ Parastatistics Fock space}
Palev shown \cite{Palev} that the  creation-annihilation superalgebra ($\ref{super}$) 
with $m$ parafer\-mio\-nic
and $n$ parabosonic degrees of freeedom is isomorphic to the
the orthosymplectic superalgebra $\mathfrak{osp}_{1+2m|2n}$. This isomorphism allows us 
 to define the parastatistics Fock space as a special 
$U(\mathfrak{osp}_{1+2m|2n})$-representation.

\begin{definition}
The representation of the universal enveloping algebra 
$U(\mathfrak{osp}_{1+2m|2n})$ built on a unique vacuum space $\rvac$ such that 
\beq
a_{i}\rvac =0  \qquad [\![ a_{i},a_{j}^{\dagger} ]\!]\rvac =p \delta_{ij} \rvac
\eeq
will be referred to as  parastatistics Fock  space $\mathcal F(m|n;p)$ of the creation-annihilation algebra (\ref{super}) with  $m$ parafermions and $n$ parabosons.
The number $p$ is called the order of the parastatistics.

The creation parastatistics algebra
 $PS(V)$ is  universal in the following sense;
the parastatistics Fock space $\mathcal F(m|n;p)$ of order $p$
 is isomorphic to the quotient 
$$
\mathcal F(m|n;p)\cong PS(V)/ M(V,p) .
$$
\end{definition}

For $p=1$ the parastatistics Fock space $\mathcal F(m|n;p)$ is the
ordinary  Fock space $\mathcal F$
of a system with $m$ fermions and $n$ bosons.


\begin{lemma} The elements $E_{ij}=\frac{1}{2}[\![ a_{i}^{\dagger},a_{j} ]\!]$
of the creation-annihilation algebra (\ref{super}) satisfy
$$
[\![ E_{ij}, E_{kl} ]\!] = E_{il} \delta_{jk} - (-1)^{(\hat{i}-\hat{j})(\hat{k}-\hat{l})}E_{jk} \delta_{il} 
$$
i.e. they close the general linear Lie superalgebra $\mathfrak{gl}_{m|n}$. 
The superspace $V$ is a fundamental representation of the
superalgebra $\mathfrak{gl}_{m|n}$, 
$ E_{ij}  a_{k}^{\dagger}= \delta_{jk}a_{i}^{\dagger}.$
\end{lemma}
 The algebra $\mathfrak{gl}_{m|n}$ can be extended to the parabolic subalgebra 
$$
\mathcal P = span \{ \, [\![ a_{i}^{\dagger},a_{j}]\!],   a_{i}, [\![ a_{i},a_{j} ]\!] \ ; 
\,\, i,j=1, \ldots, m+n    \}  $$
thus we have the chain of inclusions
$
\mathfrak{gl}_{m|n} \subset \mathcal P \subset \mathfrak{osp}_{1+2m|2n}.
$
The subalgebra $\mathcal P$ acts trivially on the vacuum space $\mathbb C \rvac$
hence the parastatistics Fock space $\mathcal F(m|n;p)$ is the induced module 
$$
\mathcal F(m|n;p)= \mbox{Ind}_{\mathcal P}^{\mathfrak{osp}_{1+2m|2n}} \mathbb C \rvac
$$
The inclusion $\mathfrak{gl}_{m|n} \subset \mathfrak{osp}_{1+2m|2n}$
implies that the space $\mathcal F(m|n;p)$
 has decomposition into irreducible representations
of $\mathfrak{gl}_{m|n}$.

The tensor powers $V^{\otimes r}$ of the vector representation $V$ 
are completely reducible $U(\mathfrak{gl}_{m|n})$-modules.
The $U(\mathfrak{gl}_{m|n})$-irreducible subrepresentations of $V^{\otimes r}$ are
indexed by Young diagrams (or partitions), i.e., in the same vein as the representations
of the symmetric group.

The roots of such a parallel of representations are in the double centralizing property
of the superalgebra action and the sign permutation
action of $\S_r$ in $\mathrm{End} (V^{\otimes r})$.
\begin{theorem}
\label{ScW}
(The Schur-Weyl duality \cite{BR}) Let the  $\mathfrak{gl}_{m|n}$-action $\rho$ on 
$V^{\otimes r}$ be
$$
\rho(X) (a^{\dagger}_{i_1} \otimes \ldots \otimes  a^{\dagger}_{i_r})
 := \sum_{k} (-1)^{p_k(X)} a^{\dagger}_{i_1} \otimes \ldots (X a^{\dagger}_{i_k} ) \ldots  \otimes  a^{\dagger}_{i_r}, \quad X \in \mathfrak{gl}_{m|n}
$$
where $p_k(X)=\hat{X} \sum_{j=k+1}^{r} \,  \hat{i}_j$. Let the sign permutation action $\sigma$
on $V^{\otimes r}$ be
$$
( a^{\dagger}_{i_1} \otimes \ldots \otimes  a^{\dagger}_{i_r})\, \sigma(\tau) :=
 \epsilon (\tau, I) \,
a^{\dagger}_{\tau^{-1}(i_1)} \otimes \ldots \otimes  a^{\dagger}_{\tau^{-1}(i_r)},
\qquad \qquad
\tau\in \S_r
$$
where $\epsilon(\tau, I) =\pm 1$ is the parity of the odd-odd (paraboson) exchanges.
The actions $\rho$ and $\sigma$ of the generators  are extended by linearity. 
The algebras  
$\sigma(\mathbb C [\S_r])$ and $\rho(U(\glmn))$ are centralizers to each other in 
$\mathrm {End}(V^{\otimes r})$
\beq
\label{sw2}
\rho(U(\glmn))=End_{\S_r}(\Vt  r)\qquad \qquad \sigma(\mathbb C[\S_r])=End_{U(\glmn)}(\Vt r) \ .
\eeq
\end{theorem}
Thus the superalgebra modules are determined from those of $\S_r$.
An irreducible $\S_r$-module $S^{\lambda}$ defines an irreducible 
$U(\glmn)$-module  $V^{\lambda}$ through  the Schur functor 
$$
\mathbb S^{\lambda}V= V^{\lambda} :=   S^{\lambda} \otimes_{\S_{r}} V^{\otimes r}
$$
where $\S_{r}$ acts on $V^{\otimes r}$  by the sign permutation action $\sigma$.
For instance the $r$-th degree of the  {\it symmetric algebra} $S(V)$
is the Schur module attached to a single row diagram with $r$ cells
$$S(V)=\oplus_{r \geq 0} \mathbb S^{r} V, \qquad \qquad 
\mathbb S^{r} V =V^{(r)} \quad ( \mathbb S^{0} V:=\mathbb C).$$ 
The map $\sigma:\mathbb C[ \S_r] \rightarrow End_{U(\mathfrak{gl}(V))}(V^{\otimes r})$ is known to be surjective and its cokernel is given by Berele and Regev Hook lemma.

\begin{theorem}( Berele and Regev \cite{BR}) The image   $\sigma(\mathbb C [\S_r])=\bigoplus_{\lambda \in \Gamma } A^{\lambda}$  of the sign representation $\sigma$ in $\mathrm {End}(V^{\otimes r})$
for the $m|n$-dimensional vector representation $V$
 is labelled by the subset $\Gamma$ of  diagrams with $r$ cells  included in a 
hook of arm-height $m$ and leg-width $n$,
$$\begin{tabular}{|cccccc}
\hline
&&&& &  $\uparrow$\\ &&&&& m \\&&&&& $\downarrow$ \\
\cline{5-6}
&&&&\!\!\!\!\!\!\!\vline \\
$\leftarrow$&&\!\!\!\!n& $\rightarrow$ &\!\!\!\!\!\!\!\vline\\
\end{tabular}$$
 $H(m,n;r)=\{ \lambda \vdash r | \lambda_j \leq n \,\, \mbox{if}  \,\,\,\,  j> m  \}$, 
\beq
\sigma(\mathbb C [\S_r]) \cong \bigoplus_{\lambda\in H(m,n;r)} S^{\lambda}\ .
\eeq
\end{theorem}
\begin{definition}
The super semistandard Young tableaux or $(m,n)$-semistandard Young tableaux  of shape $\lambda$ are the fillings of the 
Young diagram $\lambda$ with the letters of the ordered signed alphabet 
$\{ 
1, \ldots , m, \bar{1},
\ldots, \bar{n}\}$ such that 
the even indices are nondecreasing in rows and
increasing in columns whereas the odd indices are increasing in rows and nondecreasing
in columns.
\end{definition}
We denote by $H(m,n)=\bigcup_r H(m,n;r)$ all the diagrams within the hook.
Note that a $(m,n)$-semistandard Young tableau is always of shape
within the hook $H(m,n)$,   $\lambda \in  H(m,n)$.

The basis of the irreducible $\mathfrak{gl}_{m|n}$-module  $V^{\lambda}$ 
is indexed by the super semistandard Young tableau of shape $\lambda \in H(m,n)$.
As $\mathbb C[\S_r]$-$U(\mathfrak{gl}_{m|n})$-bimodule the superspace  $\Vt r$ is isomorphic to 
$$
\Vt r \cong \bigoplus_{\lambda\in H(m,n; \, r)}   S^{\lambda}\otimes V^{\lambda}  \ .
$$

The irreducible  $\mathfrak{gl}_{m|n}$-module $V^{\lambda}$ can be lifted to a
module of the  supergroup $GL(m|n)$ \cite{Fioresi}, thus $V^{\lambda}$ can also be referred to as linear supergroup $GL(V)$-module.

The Weyl theorem for the polynomial  $GL(m)$-modules
is a particular instance of the Hook theorem when $V=V_0$ is a (bosonic) vector space,
 since $H(m,0)$ are the diagrams with no more than $m$ rows. 
The basis of an  irreducible $GL(m)$-module $V^{\lambda}$ with $\lambda_{m+1}=0$
is labelled by the usual semistandard  Young  tableaux (with indices nondecreasing in rows and increasing in columns) of shape $\lambda$. 

It is worth noting
  that $H(0,n)$ are the diagrams 
within a vertical strip, i.e., with no more than $n$ columns,
 which label  the $GL(0|n)$-modules when $V$ is a fermionic space $V=V_1$.
The isomorphism $GL(0|n)\cong GL(n)$ put in correspondence
a Young tableau to its transposed.

\section{
The $U(\glmn)$-module $PS(V)$ and the $\S$-module $PS$}

\begin{lemma}
The   double supercommutator subspace $I_3(V)=[\![V,[\![V,V]\!]_{\otimes} ]\!]_{\otimes}\subset V^{\otimes 3}$   is an irreducible Schur module
$$V^{(2,1)}=I_3(V)
= e \, \mathbb C[\S_3] \otimes_{\S_3}  V^{\otimes 3} $$ 
arising as the Schur functor image of the $\S_3$-module $I(3)= e \, \mathbb C[\S_3]\cong S^{(2,1)}$, where $e$ stands for the
Eulerian idempotent \cite{eulerien}
$$   e=\frac{1}{3}\left(123 -\frac{1}{2} (231+213+132+312)  + 321\right).$$
\end{lemma}
\begin{proof}
The cyclic  permutation of $I_3(V)$ vanishes 
due to  the super Jacobi idenity
$$
[\![x,[\![y,z]\!] ]\!] + (-1)^{\hat{x}\hat{y}+\hat{x}\hat{z}}[\![y,[\![z,x]\!] ]\!] +(-1)^{\hat{x}\hat{z}+\hat{y}\hat{z}}[\![z,[\![x,y]\!] ]\!]= 0 \ , 
\qquad x,y,z \in V 
$$
thus $I_3(V)\cap V^{(1^3)}=0=I_3(V)\cap V^{(3)}$ and, counting the dimensions,
we conclude that $I_3(V)=V^{(2,1)}$. For the $\S_3$-representation $I(3)$ the  Jacobi identity 
implies $$I(3)=\mathrm{Ind}_{\mathbb Z_3}^{\S_3} 1\!\!1 \ . $$
 The rest is a direct calculation.
\end{proof}

\begin{theorem} Let $V$ be the $m|n$-dimensional super space.
In the  decomposition of the $U(\glmn)$-module $PS(V)$ into irreducibles 
each  
$V^{\lambda}$, $\lambda\in H(m,n)$ appears once and exactly once
$
PS(V) \cong  \bigoplus_{\lambda \in H(m,n)} V^{\lambda}.
$
\label{once}
\end{theorem}
\begin{proof} Let us consider first the case of an  $m$-dimensional even space $V=V_0$.
The left hand side of the Schur formula 
$$
\prod_{i=1}^m \frac{1}{1-x_{i}} 
\prod_{1\leq i<j\leq m} \frac{1}{1-x_{i}x_{j}} =
\sum_{\lambda} s_{\lambda}(x)
$$
is the   character of the $U(\glm)$-module $PS(V)\cong S(V)\otimes S([V,V])$ in view of the  Poincar\'e-Birkhoff-Witt theorem.
Then the sum of the Schur polynomials $s_{\lambda}(x)$ 
(which are characters of the irreducible $U(\glm)$-modules $V^{\lambda}$)
 on the right hand side implies  
 $
    PS(V) \cong  \bigoplus_{\lambda} V^{\lambda}= \bigoplus_{\lambda\in H(m,0)} V^{\lambda}
 $ for $    V=V_0
$
where the sum  on $\lambda$ runs on the Young diagrams  with no more than $m$ rows,
 $ \lambda_{m+1}=0$. Thus all nontrivial $U(\glm)$-modules modules are present in 
$PS(V_0)$.

\begin{lemma}Let us have ${\S}=\bigoplus_{r\geq 0} \S_r$.
 The decomposition of the $\S$-module
    $PS= \bigoplus_{r\geq 0} PS(r)$ contains each irreducible finite
    dimensional $\mathbb C[\S_{r}]$-module $S^{\lambda}$, $r\geq 0$,
     exactly once
     $
PS = \bigoplus_{\lambda} S^{\lambda}.   
$
   \end{lemma}
   \noindent
    {\bf Proof of the lemma.}
    We have $PS(V_0) = \bigoplus_{r\geq 0} PS_r(V_0)$. Let us denote by $PS(r)$
 the multilinear part of $PS(V_0)$   of the $r$-homogeneous Schur functor 
$PS_{r}(V_0)$ for even space $V_0$  of dimension $r$.
The space $PS(r)$ is a reducible $\S_r$-module and from the decomposition of $PS(V_0)$
follows
$
    PS(r) \cong 
    \bigoplus_{\lambda \vdash r} S^{\lambda}.
    $
The statement of the lemma 
follows by induction on the dimension $r$.

Now let us take $V$ to be a $m|n$-dimensional space.
 It is enough to apply the Schur functor $PS$ to  the superspace $V$.
 The nontrivial $U(\glmn)$-modules $V^{\lambda}$ are labelled by Young diagrams within
the $(m,n)$-hook 
 and all these appear exactly once.
Since $V^{\lambda} \equiv 0$ iff $\lambda \notin H(m,n)$ we get
$
PS(V) \cong  \bigoplus_{\lambda \in H(m,n)} V^{\lambda}.
$
\end{proof}

\begin{corollary}
\label{hsf}
The hook generalization of the Schur  identity reads
\beq
\label{ident}
 \frac{ \prod_{i<j \,,\,\hat{i}\neq \hat{j}} (1+x_{i}x_{j})}
 { \prod_{i}(1-x_{i}) \prod_{i<j \,,\,\hat{i}=\hat{j}}{(1-x_{i}x_{j})}  }
=\sum_{\lambda} hs_{\lambda} (x_1,  \ldots, x_{m+n})
\eeq
where $hs_{\lambda}(x)$ stands for the Hook Schur function of $m$ even and
$n$ odd variables.
\end{corollary}
\begin{proof}
The $U(\glmn)$ character of the  Schur module $V^{\lambda}$ 
, $\lambda \in H(m,n)$ of the $m|n$-dimensional superspace  is the 
hook Schur function 
\cite{BR}
$$
hs_{\lambda}(x_1,  \ldots, x_{m+n})
= \sum_{\mu \subset \lambda} s_{\mu}(x_1, \ldots, x_m)
 s_{\lambda'/ \mu'} ( x_{m+1}, \ldots, x_{m+n}).
$$
where $\lambda'$ stands for the transposed Young diagram of $\lambda$. The 
hook Schur functions have also a combinatorial definition
$$
hs_{\lambda}(x)= \sum_{T\in SSYT(\lambda)} x^{wt(T)}  \qquad \qquad \lambda \in H(m,n)
$$
where the sum runs over all $(m,n)$-semistandard Young tableaux of shape $\lambda$.
One has $hs_{\lambda}(x)=0$ iff $\lambda \notin H(m,n)$.

The algebra $PS(V)$ is the universal enveloping algebra of the $2$-nilpotent Lie superalgebra (\ref{PBW}), hence the Poincar\'e-Bikhoff-Witt theorem for superalgebras yields
$$
PS(V)\cong \mathbb S(V\oplus \mathbb S^{1^2}V)\cong S(V_0)\otimes \Lambda(V_1)\otimes S(\Lambda^2(V_0))\otimes S( S^2(V_1))
\otimes \Lambda(V_0 \wedge V_1) \ .
$$
Therefore the character $\chi_{PS(V)}(x)$ of the $U(\glmn)$-module $PS(V)$ reads
$$
\chi_{PS(V)}(x) =  \frac{\prod_{i, \hat{i}=1}(1+ x_{i})}{\prod_{i, \hat{i}=0} (1-x_i)} 
\frac{\prod_{i<j \,,\,\hat{i}\neq\hat{j}} (1+ x_i x_j)}{\prod_{i<j \,,\,\hat{i}=\hat{j}}(1-x_i x_j )\prod_{i: \hat{i}=1}(1-x_i^2)}
$$
which equals to the LHS of the eq. (\ref{ident}). On the other hand
 the character $\chi_{PS(V)}$ is  a sum of the
characters $hs_{\lambda}$ of the irreducible components $V^{\lambda}$ which
ends the proof of the identity (\ref{ident}).
\end{proof}

\section{Parastatistics Fock spaces revisited.} 
Parastatistics Fock space 
 $\mathcal F(m|n;p)$ of order $p$ is  the quotient of the $PS$-algebra
$$
 \mathcal F(m|n;p)= PS(V)/ M(V,p) \ ,
$$
where the graded ideal $M(V,p)$ is generated by the irreducible Schur module 
$V^{(p)}=\mathbb S^{p+1} V \subset
  \Vt{ p+1}$,
$$
M(V,p) = (\mathbb S^{p+1}V)\ .
$$
For parafermions $V=V_0$ and the space  $\mathbb S^{p+1}V$ is the ordinary $p+1$-symmetrizer 
corresponding to the one row diagram
$$\underbrace{\begin{scriptsize}\tiny
\ba{|c|c|c|}
\hline \, \, & \,\, & \, \, \\
\hline
\ea
\ldots
\ba{|c|c|c|}
\hline \, \, & \, \, & \, \, \\
\hline
\ea 
\end{scriptsize}}_{p+1}$$
 therefore in the Fock space $ \mathcal F(m|0;p)$ no more than $p$ identical paraparticles can occupy one state (tableau). 
Hence in general factoring by $\mathbb S^{p+1}V$ the parastatistics algebra $PS(V)$ is the  superfication of the Fermi  exculsion principle.

\begin{definition}
The $\S$-submodule  isomorphic to
the factor of the $\S$-module $PS$  
\beq \label{psp}
PS^{(p)}  \cong PS / (S^{({p}+1)})
\eeq
 such that
$\mathcal F(m|n;{p}) \cong PS^{(p)} (V)=\bigoplus_{r\geq 0} PS^{(p)}(r) \otimes_{\S_r} \Vt r 
$
will be referred to as $p$-restricted parastatistics $\S$-module 
$$PS^{(p)}= \bigoplus_{r\geq 0}PS^{(p)}(r) \qquad \qquad
PS^{(p)}(r) \subset \mathbb C [\S_r] \ .$$
\end{definition}

The decomposition of the $p$-restricted $\bf \S$-module
    $PS^{(p)}$ contains 
once and  exactly once
      each irreducible finite
    dimensional $\mathbb C[\S_{r}]$-module $S^{\lambda}$ 
such that its partition $\lambda$ is restricted by 
${p}\geq \lambda_1 \geq \ldots \lambda_r \geq 0$, 
   \beq
PS^{(p)} = \bigoplus_{\lambda \, : \lambda_1 \leq {p}} S^{\lambda}  \qquad \qquad  
\eeq
which means that we eliminate from $PS$ all the irreducibles
 with diagrams with more than $p$ columns.

Let us denote by $F_0$ the set of the self-conjugated partitions $\eta =\eta'$, i.e., the partitions $\eta$ with  Frobenius notation 
\beq
\label{frob}
\eta = (\alpha_1 ,\, \alpha_2 , \ldots ,\alpha_r \, | \alpha_1,\alpha_2, \ldots, \alpha_r ),
\qquad \alpha_1> \alpha_2> \ldots>\alpha_r \geq 0 \ ,
\eeq
and  by $F_p$ of the set of the partitions with $p$-augmented arms
$$\eta_{(p)}=(\alpha_1 +p,\, \alpha_2 +p, \ldots ,\alpha_r +p\, | \alpha_1,\alpha_2, \ldots, \alpha_r ),
\qquad \alpha_1> \ldots>\alpha_r \geq 0 \ .
$$
Thus $\eta_{(p)}\in F_p$ when $\eta\in F_0$.
One has $S^{\eta_{(p)}} \subset (S^{(p+1)})$.

\begin{corollary} 
The parastatistics Fock space $\mathcal F(m|n;p)$ of order $p$ as a $U(\glmn)$-module
 is isomorphic to the sum of Schur modules 
\beq
\mathcal F(m|n;{p}) \cong  
\bigoplus_{
 \lambda \in H^{(p)}(m,n)}
 V^{\lambda} 
\eeq
where $H^{(p)}(m,n)$ stands for the set of hook partitions $H(m,n)$ with no more than $p$ columns, $\lambda_1\leq p$.
The character formula for $\mathcal F(m|n;p)$ yields the identity
\beq
\label{identp}
 \frac{ \sum_{\eta \in F_0} (-1)^{\frac{1}{2}(|\eta| + r)} hs_{\eta_{(p)}}(x) \prod_{i<j \,,\,\hat{i}\neq \hat{j}} (1+x_{i}x_{j})}
 { \prod_{i}(1-x_{i}) \prod_{i<j \,,\,\hat{i}=\hat{j}}{(1-x_{i}x_{j})}  }
=\sum_{\lambda:\, \lambda_1\leq { p}} hs_{\lambda}(x) \ .
\eeq
\end{corollary}
{\bf Proof.}
For an even space $V=V_0$ (i.e., parafermions only) the $U(\mathfrak{ gl}(V))$-character 
$\chi_{PS^{(p)}(V)}$
of the parastatistics Fock space $PS^{(p)}(V)\cong \mathcal F(m|0,{p}) $ reads
\beq \label{fp}
\chi_{PS^{(p)}(V)}(x)= \sum_{\lambda:\, \lambda_1\leq {p}} s_{\lambda}(x)=
 \frac{
\sum_{\eta \in F_0} (-1)^{\frac{1}{2}( |\eta|+ r)} s_{\eta_{(p)}}(x)}
{\prod_{i} {(1-x_{i})} 
\prod_{i<j}{(1-x_{i}x_{j}} )} 
 \ .
\eeq
Here the sum over the $p$-restricted Young diagrams is  the character of
$PS^{(p)}(V)$ and the sum over the diagrams from $E_{(p)}$ (\ref{frob}) is the character of
the factor $PS(V)/M(V,p)$.  In the works on parastatistics \cite{LSV,SVdJ} 
the proof of the identity (\ref{fp}) is  attributed to R. King\footnote{We are grateful to Ron King for sending us his proof of the identity (\ref{fp}).}.
For our proof of the character identity  see the appendix.


Now let $V$ be a superspace  of dimension $m|n$. 
From the functoriality of the construction of the Fock space $\mathcal F (m|n;p)$
as a submodule and factor-module of $PS(V)$ 
 $$\mathcal F (m|n;p)=PS(V)/ M(V,p) \cong PS^{(p)} (V)$$
it follows the identity between the Hook Schur functions
 \beq
\label{identp1}
\frac{ \sum_{\eta \in F_0} (-1)^{\frac{1}{2}(|\eta| + r)} hs_{\eta_{(p)}}(x) }
{ \sum_{\eta \in F_0} (-1)^{\frac{1}{2}(|\eta| + r)} hs_{\eta}(x) }=
\sum_{\lambda:\, \lambda_1\leq { p}} hs_{\lambda}(x) 
 \ .
\eeq
 Then the formula  for the $U(\glmn)$-character $\chi_{{PS}(V)}$
 (\ref{ident})
 implies
the identity (\ref{identp}). $\Box$

\begin{remark}  The $N$-complexes \cite{D-V} arise in a natural way in the  approach of Dubois-Violette and Marc Henneaux \cite{D-VH1,D-VH2}
(see also the Marc Henneaux's contribution to this volume) to the higher 
gauge spin fields. In the study of the gauge $S$-spin fields (with   $S\geq 1$)
the $(S+1)$-complexes are involved. The  degrees of these complexes are spaces 
of tensors with Young-symmetry type with the  constraint that the  row lengths of the corresponding Young diagrams  are no longer than the spin $S$. The components of the tensor fields on  $\mathbb R^D$
are labelled by different Young tableaux (with entries from $1$ to $D$)
 and the total space of the $(S+1)$-complex has 
the same structure as the
the parafermionic Fock space ${\mathcal F}(D| 0, S)$ of parastatistical order $S$.
\end{remark}

\section{Deformed parastatistics and its para-Fock Space}

The parastatistics algebras (\ref{super}) of creation and annihilation operators allow for  $q$-deformations as introduced by Palev \cite{Palev2}. The idea is to replace the universal enveloping algebra (UEA)  $U(\mathfrak{osp}_{1+2m|2n})$ by the quantum UEA $U_q(\mathfrak{osp}_{1+2m|2n})$ written in an alternative form, with a system of relations between generators corresponding to
the parastatistics creation and annihilation operators.
 We are going to describe the deformation $\PS(V)$ of 
the creation parastatistics algebra $PS(V)$. The space $\PS(V)$ is naturally  a $U_q(\glmn)$-module and instead of working with the 
$U_q(\mathfrak{osp}_{1+2m|2n})$ relations we choose another approach based on the $q$-Schur modules and the Hecke algebra.
Our aim is to extract from $\PS(V)$ a combinatorial algebra having as elements the super semistandard Young tableaux.
\subsection{Hecke algebra.}
The {\it Hecke algebra } $H_{r}(q)$ is the associative algebra generated by $g_{1}, \ldots , g_{r-1}$ with the 
    relations
     \beq
\begin{array}{rcll}
      g_{i}g_{i+1}g_{i} & = & g_{i+1}g_{i}g_{i+1}  & \quad i=1,\ldots, r-1
    \\
      g_{i}g_{j} & = & g_{j}g_{i}  & \quad  |i-j|\geq 2 \\
      g_{i}^{2} &= & {1} + (q - q^{-1})g_{i}  &\quad i=1,\ldots, r-1
     \end{array}
 \label{Hck}
\eeq
The specialization $q=1$ yields the Coxeter relations of the symmetric 
  group $\S_{r}$ generated by the elementary transpositions
     $s_{i}=(i\, i+1)$ for $i=1, \ldots, r-1$.

 The elements of  $\hr$ are indexed by  permutations in
 $\sigma \in \S_r$,
$T_{\sigma}\in \hr $ in the following way. 
Let $s_{i_1} \ldots s_{i_k}$ be the reduced word 
of the permutation $\sigma$, then 
$$T_{\sigma}:= T_{s_{i_1}} \ldots  T_{s_{i_k}}
 \qquad \sigma=s_{i_1} \ldots s_{i_k}$$
where $T_{s_{i}}=g_i$ and the Coxeter relations (\ref{Hck}) are equivalent to the relations
\beqa
T_{\rho}T_{\sigma} &=&T_{\rho \sigma} \qquad \qquad  \qquad \mbox{when}\quad |\rho \sigma|=|\rho| +|\sigma|\\
T_{s_{i}}^2&=&1 +(q - q^{-1})T_{s_{i}}
\eeqa

For $q$ generic  the Hecke algebra $H_r(q)$ is isomorphic to the group algebra $\mathbb C[\S_r]$.
The irreducible $H_r(q)$-modules  ${\mathcal H}^{\lambda}$ are indexed
by Young diagrams with $r$ boxes $\lambda\vdash r$, i.e., 
in the same manner as the irreducible 
$\S_r$-modules.  

\subsection{Parastatistics Hecke ideal}
We now consider the  $H_3(q)$-module $\I(3) \cong{\mathcal H}^{(2,1)}$ which is a
deformation of the $\S_3$-module  $I(3) = e \mathbb C [\S_3] \cong S^{(2,1)}$.
To this end we find an idempotent  $e(q)\in H_3(q)$ which is a deformation of the Eulerian idempotent $e$, in the sense that $e(1)=e$.


Let us  denote by $\omega$ the maximal element in $H_3(q)$, $\omega=g_1g_2g_1$.
\begin{lemma}
The  equation for the idempotent $e(q)\in H_3(q)$
\beq \omega e(q)=e(q)
\label{om} \eeq fixes  $e(q)$ completely when $q\neq \pm 1$.
The unique solution is given by the expression
\beqa 
  \label{idemp} 
e(q)&:=&
\frac{1}{[3]}\left(T_{123} -\frac{1}{2} (T_{231}+ T_{213}+ T_{132}+T_{312}) 
+ T_{ 321}\right) \nonumber \\
&+&\frac{q - q}{2[3]}^{-1} \left( T_{213}- T_{312} - T_{231} + T_{132}\right).
\eeqa
\end{lemma}
\begin{proof} The statement can be checked by direct calculation in
the Hecke algebra $H_3(q)$ taking into account 
$T_{123} = 1\!\!\!1, \ T_{213}=g_1, \ T_{132}=g_2, \ T_{213} =g_1 g_2, \ T_{312}=g_2 g_1, \ T_{321}=g_1g_2g_1=\omega$. 
\end{proof}
Thus the symmetry $\omega e =e$ which holds true for the Eulerian idempotent $e$ in the group algebra $\mathbb C [\S_3 ]$ is preserved during the deformation.

Let us define $\I(3)$  to be the right $H_3(q)$-ideal $\I(3)= e(q) H_3(q)$.
By construction $\I(3)$ is an $H_3(q)$-ideal such that its  ''classical'' limit $q\rightarrow 1$, i.e., its specialization to the $\mathbb C[\S_3]$-module is $I(3)$.  
It is spanned by two elements ${\bf \Gamma}^{12}_3$ and  ${\bf \Gamma}^{13}_2$ 
\beq
\I(3)  = \mathbb C(q){\bf \Gamma}^{12}_3 \oplus \mathbb C(q) {\bf \Gamma}^{13}_2 
\eeq
which can be chosen to be
\beqa
{ \bf \Gamma}^{13}_2&=&q (T_{213} - T_{231})+ T_{123}- T_{132} -T_{231}+ T_{321} +q^{-1}(T_{312}- T_{132} ) \ , \\
{\bf \Gamma}^{12}_3&=&q (T_{132}- T_{312}) + T_{123} - T_{213} - T_{312} +T_{321} + q^{-1}(T_{231} -T_{213})\ .
\eeqa
Written in terms  of the $H_3(q)$-generators $g_1$ and $g_2$ the basis of $\I(3)$ looks like
\beqa 
{\bf \Gamma}^{13}_2&=&1\!\!1 + q g_1 -(1+ q^{-1})g_2 -(1 + q )g_2g_1 + q^{-1} g_1g_2 + g_2g_1g_2 \ ,
\nn
{\bf \Gamma}^{12}_3&=&1\!\!1 + q g_2 -(1+ q^{-1})g_1 -(1 + q )g_1g_2 + q^{-1} g_2g_1 + g_1g_2g_1 \ ,
\nonumber
\eeqa
and the $H_3(q)$-action is determined by linear continuation of the actions 
\beq
\ba{lcccclcc}
{\bf \Gamma}^{12}_3 g_1 &= &-q^{-1} {\bf \Gamma}^{12}_3- {\bf \Gamma}^{13}_2 
&\,\, &\,\, & {\bf \Gamma}^{13}_2 g_1 &=& q {\bf \Gamma}^{13}_2 \\
{\bf \Gamma}^{12}_3 g_2 &=& q \, {\bf \Gamma}^{12}_3 
&\,\, &\,\, &
{\bf \Gamma}^{13}_2 g_2 &= &-q^{-1} {\bf \Gamma}^{13}_2- {\bf \Gamma}^{12}_3 \ .
\ea 
\eeq

\subsection{Quantum Schur-Weyl Duality}
We are now going to sketch how  the Hecke algebra  is related to the deformation 
 $U_q(\glmn)$ of UEA $U(\glmn)$ providing a generalization of the Schur-Weyl duality.
For the defining relations of the quantum UEA $U_q(\glmn)$ we send the reader to \cite{BKK}.

 We need some preliminaries on the $R$-matrix.
 \begin{lemma}  \label{Rmatr}
 Let $V$ be a superspace of dimension $m|n$ over the field $K(q)$. The linear transformation
$\hat{R}\in End( V\otimes V)$ given by
the $(m+n)^2\times(m+n)^2$ matrix 
\beq
\label{superR}
\hat{R}^{i \, j }_{k \, l }= 
(-1)^{\hat{i} \hat{j}} q^{(-1)^{\hat{i}} \delta_{ij} }\,   \delta^{i}_{l}\delta^{j}_{k}
+ (q - q^{-1}) \, \theta_{ji}\, \delta^{i}_{k}\delta^{j}_{l}
\eeq
is an $R$-matrix of the quantum linear supergroup $GL_q(m|n)$, i.e., $\hat{R}$ satisfies \\
i) the Yang-Baxter equation 
$$
 \hat{R}_{1}\hat{R}_{2}\hat{R}_{1}=\hat{R}_{2}\hat{R}_{1}\hat{R}_{2}\, 
\qquad \quad (\hat{R}_{1}= \hat{R}\otimes 1\!\! 1 \quad  \hat{R}_{2}=1\!\! 1\otimes \hat{R}) \ ,
$$
ii) the Hecke relation
$$
\hat{R}^2=1 \!\! 1 +(q- q^{-1}) \hat{R} \ ,
$$
having two eigenvalues $\pm q^{\pm 1}$ appearing with multiplicities $\frac{m(m\pm 1)}{2}+ \frac{n(n\mp1 )}{2}+mn$. The discrete step function $\theta_{ij}$ is $0$ when $i<j$ and
$1$ when $i\geq j$.
\end{lemma}

\subsection{Sign permutation action of the Hecke algeba $H_r(q)$.} Let us consider the
left  action $\sigma_q: \Vt r \rightarrow \Vt r $ defined for the $H_r(q)$ generators by
\beq
\label{qsign}
\pi_q(g_s)\,a^{\dagger}_{i_1} \ldots a^{\dagger}_{i_s} a^{\dagger}_{i_{s+1}}\ldots a^{\dagger}_{i_r}=
\sum_{j_s , j_{s+1}}a^{\dagger}_{i_1} \ldots a^{\dagger}_{j_s}a^{\dagger}_{j_{s+1}}\ldots a^{\dagger}_{i_r} \hat{R}^{j_s j_{s+1}}_{i_s i_{s+1}} 
 \quad  s\leq r-1
\eeq
and extended by linearity. This action is indeed a  $H_r(q)$-representation
by virtue of the Lemma \ref{Rmatr} and  it will be referred to as
{\it sign permutation action of} $H_r(q)$.
In more details the action (\ref{qsign}) reads
$$
\pi_q(g_s) \,a^{\dagger}_{I} =
\left\{
\ba{lcc}
(-1)^{\hat{i}_s} q^{(-1)^{\hat{i}_s}} a^{\dagger}_{I} &&i_s=i_{s+1}\\ 
(-1)^{\hat{i}_s \hat{i}_{s+1} } a^{\dagger}_{ \sigma_s^{-1} (I)}+ (q-q^{-1})e_{I} & &i_s< i_{s+1} \\
(-1)^{\hat{i}_s \hat{i}_{s+1} } a^{\dagger}_{ \sigma_s^{-1} (I)} && i_s> i_{s+1} 
\ea \right.
$$
where the  (multi)index permutation $\sigma_s^{-1}(I)=i_1 \ldots i_{s+1} i_{s} \ldots i_r$.
In the limit $q \rightarrow 1$ one retrieves the sign permutation action  of the symmetric group,
\[
u\otimes v \, \hat{R}\,|_{q\rightarrow 1} = (-1)^{\hat{u}\hat{v}} v \otimes u \ .
\]


\begin{theorem} (quantum Schur-Weyl duality \cite{Mitsuhashi})
\label{qSWglmn}
The sign permutation action $\pi_q$
of the Hecke algebra $\hr$ and  the action $\rho$ of the quantum UEA $U_q(\glmn)$ 
on $\Vt r$ are centralizers
to each other
\beq
\label{sw3}
\rho(U_q(\glmn))=End_{\hr}(\Vt  r) \ ,\qquad \qquad \pi_q(\hr)=End_{U_q(\glmn)}(\Vt r) \ . 
\eeq
\end{theorem}
The quantum version of the Schur-Weyl duality between the quantum UEA  $U_q(\glm)$-action on the tensor power of the vector representation $V^{\otimes r}$ and the permutation action of the Hecke algebra $\hr$ is due to Jimbo \cite{Jimbo2},
whereas its super-counterpart given by  Theorem \ref{qSWglmn} was done 
 by Mitsuhashi \cite{Mitsuhashi}.

 \subsection{
The $q$-Schur functor.}
The  quantum Schur-Weyl duality stated in Theorem \ref{qSWglmn} allows to build the representations
of the quantum UEA $U_q(\glmn)$ from the Hecke modules in the same fashion as the 
representations of $U(\glmn)$ (and $GL(m|n)$) are built from $\S$-modules.

Let us have  $H$-module $\mathcal M$ which is a family of right $\hr$-modules $\mathcal M(r)$, $r\geq 0$.
Its associated $q$-Schur functor $\mathcal M: gVect \rightarrow gVect$ is defined as
\beq
{\mathcal M}(V):= \bigoplus_{r\geq 0} \mathcal M(r) \otimes_{\hr} V^{\otimes r}
\eeq
where $V$ is a superspace  over the field $K(q)$ and the $\hr$-action
on $\Vt r$ is the sign permutation action  $\pi_q$(\ref{qsign}).
The image $\mathcal M(V)$ carries the structure of $U_q(\glmn)$-module.
The homogeneous components of 
 $\mathcal M(V)$ are denoted by
\beq
\label{mr}
 \mathcal M_{r}(V) :=  \mathcal M(r)  \otimes_{\hr} V^{\otimes r} \ , \qquad  \qquad \mathcal M(V)=\bigoplus_{r\geq 0} \mathcal M_{r}(V) \ ,
\nonumber 
\eeq
 and their irreducible $U_q(\glmn)$-submodules are denoted by
\beq
\label{irh}
V^{\lambda} =
\H^{\lambda} \otimes_{\hr} \Vt r \ , \qquad \qquad \lambda \in H(m,n) \ .
\eeq

\begin{definition}
The {\it braided parastatistics superalgebra} $\PS(V)$ is the factor algebra of the tensor algebra of $V$ by the ideal $\mathcal I(V)$
$$
\PS(V) = T(V)/ \mathcal I(V)
$$
where  $\mathcal I(V) = \bigoplus_{r \geq 3} \mathcal I_{r}(V)$ is the twosided ideal
 generated by $ \I_{3}(V)$
\beq
\mathcal I_{r}(V)= \sum_{i+j+3=r} V^{\otimes i}\otimes  
 \mathcal I_{3}(V)  \otimes V^{\otimes j}, \qquad r\geq 3.
\eeq
and  $\I_{3}(V)$ stands for 
the image of the right $H_{3}(q)$-module $\I(3)=e(q) H_q(3)$   by the $q$-Schur functor
 \[
 \I_{3}(V) =  \I(3) \otimes_{H_{3}(q)} V^{\otimes 3} \, .
 \]
\end{definition}

\begin{proposition} (\cite{LP})
     Let $a_{i}^{\dagger}$ be a basis of the $m|n$-dimensional superspace 
     $V=
     \oplus 
 \mathbb C (q) a_{i}^{\dagger}$. 
 The superspace $\I_{3}(V)\cong V^{(2,1)}$ is an irreducible $U_q(\glmn)$-module 
\beq \label{I3}
\I_{3}(V)= \bigoplus_{{}^{i_1 i_2}_{i_3} }
 \mathbb C(q)\,\, \Gamma^{i_1 i_2}_{i_3} 
\eeq
where the sum runs over all $(m,n)$-Semistandard Young Tableaux of shape $(2,1)$
and the spanning elements (chosen to be polynomial in $q^{-1}$)  read
$$
 \begin{array}{rlcllcl}
   \Gamma^{i_1 i_3}_{i_2} :=& \quad    
[\![ a^{\dagger}_{i_2}, [\![ a^{\dagger}_{i_3},a^{\dagger}_{i_1} ]\!] ]\!]_{q^{-2}} 
   &+&q^{-1}&[\![ a^{\dagger}_{i_3}, [\![a^{\dagger}_{i_1}   , a^{\dagger}_{i_2}]\!]]\!] &\quad &  i_1<i_2< i_3  \ ,
\\ [4pt]
    \Gamma^{i_1 i_2}_{i_3} :=&\quad 
  [\![ [\![ a^{\dagger}_{i_3},a^{\dagger}_{i_1} ]\!],  a^{\dagger}_{i_2}]\!]_{{q}^{-2}}
   &+&{q^{-1}}& [\![  [\![a^{\dagger}_{i_2},  a^{\dagger}_{i_3}]\!], a^{\dagger}_{i_1}]\!]
   &\quad & i_1< i_2<i_3 \ , \\[4pt]
   \Gamma^{i_1 i_2}_{i_2} :=&\quad 
[\![ [\![ a^{\dagger}_{i_1},a^{\dagger}_{i_2} ]\!], a^{\dagger}_{i_2} ]\!]
_{q^{-1}}
 &&&
   &\quad & i_1< i_2 \ , \quad  \hat{i}_2=1 \ , \\[4pt]
   \Gamma^{i_1 i_2}_{i_2} :=&\quad 
[\![ a^{\dagger}_{i_2},[\![ a^{\dagger}_{i_1},a^{\dagger}_{i_2} ]\!] ]\!]
_{q^{-1}}
 &&&
   &\quad & i_1< i_2 \ , \quad  \hat{i}_2=0 \ , \\[4pt]
    \Gamma^{i_2 i_2}_{i_3} :=&\quad 
    [\![  a^{\dagger}_{i_2} ,[\![ a^{\dagger}_{i_2},a^{\dagger}_{i_3} ]\!] ]\!]_{q^{-1}}
    &&&
   &\quad & i_2< i_3 \ , \quad \hat{i}_2=1 \ ,\\[4pt]
   \Gamma^{i_2 i_3}_{i_2} :=&\quad 
    [\![  [\![ a^{\dagger}_{i_2},a^{\dagger}_{i_3} ]\!], a^{\dagger}_{i_2}  ]\!]_{q^{-1}}
    &&&
   &\quad & i_2< i_3 \ , \quad \hat{i}_2=0 \ .
  \end{array}      
  $$
\end{proposition}
\begin{proof}
The Schur functor attached to the $H_3(q)$-module 
$\I(3)= \mathbb C(q){ \bf \Gamma}^{1 2}_{3} \oplus \mathbb C(q){ \bf \Gamma}^{1 3}_{2}$ 
is the $U_q(\glmn)$-module $\I_3(V)$ of the relations of the algebra $\PS(V)$.
 The direct calculation  of the quantities 
${\bf \Gamma}^{1 2}_{3} \otimes_{H_3(q)} (a^{\dagger}_{i}\otimes a^{\dagger}_{j}\otimes a^{\dagger}_{k})$
and ${\bf \Gamma}^{1 3}_{2} \otimes_{H_3(q)} 
(a^{\dagger}_{i}\otimes a^{\dagger}_{j}\otimes a^{\dagger}_{k})$
 yields elements which are either proportional (with coefficients in $\mathbb C(q)$) to 
some  $\Gamma^{i j}_{k}$ from (\ref{I3}) or zero.
By construction these elements span $\I_{3}(V)$.
\end{proof}

\section{Plactic monoid.} 
Let us consider the free monoid of words written in the alphabet $A$, 
the multiplication being the juxtaposition of words.
The plactic monoid \cite{LS} is the set of  the equivalence classes in the free monoid for the equivalence defined as the words with $P$-equivalent tableaux in the Robinson-Schensted
 $(P,Q)$-correspondence. The $P$-equivalence coincides with the equivalence with respect to the Knuth relations
 $$
    \ba{cc}
 x z y =  z x y  \qquad  
    & x\leq y<z 
\\ [4pt] 
      y  x z =    y z x  \qquad  &
   x<y\leq z  
   \ea
    $$ 
with $x,y,z \in A$. The classes of equivalent words in the plactic monoid
are in bijection with the Semistandard Young Tableaux 
with entries from the alphabet $A$. The algebra of the plactic monoid
will be denoted as $Plac(A)$.

On the other hand in view of Theorem \ref{once} and Corollary \ref{hsf} the states in the parastatistics algebra $PS(V)$ with 
$m$ parafermi degrees
of freedom are in bijection 
with the Semistandard Young Tableau with entries from $\{ 1 , \ldots , m\}$, that is,
the set of indices of the vector space $V=V_0$.

This parallel suggests an interrelation between  the
algebra $PS(V)$  and  the plactic algebra $Plac(V)$. 
Surprisingly in revealing this interrelation
the quantum UEAs play a key role \cite{DJM}.

\section{Plactic Superalgebra and the Parastatistics Algebra}

Let $K$ be the subring of rational functions 
without a pole at $q^{-1}$, $K\subset \mathbb C(q)$. 
By the evaluation map $f(q^{-1}) \mapsto f(0)$ we have an isomorphism
$ K / q^{-1} K \cong \mathbb C$.

\begin{definition}(Kashiwara \cite{Kashiwara}) Let $W$ be a $\mathbb C(q)$-vector space.
The local base of $W$ at $q^{-1}=0$  is the pair $(L,B)$ where 
$L$ is a free $K$-module and $B$ is a base of the $\mathbb C$-vector space 
$L/ q^{-1} L$.
\end{definition}
Let us denote by $\Gamma$ the basis of the space $\I_3(V)$, $$\Gamma = \{\quad  \Gamma^{i_1 i_2}_{i_3} \quad  | \quad {}^{i_1 i_2}_{i_3} \quad  \mbox{ is a $(m,n)$-semistandard hook tableau}\}$$
and by $\gamma$ the base of $\Gamma/ q^{-1} \Gamma$.

By applying the evaluation map one gets the following
\begin{corollary} 
The base $\gamma$ is given by the elements
$$
    \ba{cccc}
   a^{\dagger}_{i_1}a^{\dagger}_{i_3}  a^{\dagger}_{i_2} - 
(-1)^{\hat{i}_1\hat{i}_3} a^{\dagger}_{i_3} a^{\dagger}_{i_1} a^{\dagger}_{i_2} \, ,
    & (i_1\leq i_2< i_3 \, \mbox{,} \,\, \hat{i}_2=0) &\mbox{or}&  (i_1< i_2\leq i_3 \,\mbox{,} \,\, \hat{i}_2=1)\\ [4pt] 
        a^{\dagger}_{i_2}  a^{\dagger}_{i_1}a^{\dagger}_{i_3} -  (-1)^{\hat{i}_1\hat{i}_3}  a^{\dagger}_{i_2}a^{\dagger}_{i_3} a^{\dagger}_{i_1} \, , &
   (i_1<i_2\leq i_3 \, \mbox{,} \,\, \hat{i}_2=0) &\mbox{or}&  (i_1\leq i_2<i_3 \,\mbox{,} \,\, \hat{i}_2=1)
    \ea
    $$  
 The  $U_{q}(\glmn )$-module $\I_{3}(V)$ has
a local basis $(\Gamma, \gamma)$ at the point $q^{-1}=0$.
   \end{corollary}
     The algebra $\PS(V)$ at the point $q^{-1}=0$ with relations
     $$
    \ba{cccc}
 x z y =  (-1)^{\hat{x}\hat{z}} z x y \, , \qquad  
    & (x\leq y<z \, \mbox{,} \,\, \hat{y}=0) &\mbox{or}&  (x< y\leq z \,\mbox{,} \,\, \hat{y}=1)\\ [4pt] 
      y  x z =  (-1)^{\hat{x}\hat{z}}   y z x \, , \qquad  &
   (x<y\leq z \, \mbox{,} \,\, \hat{y}=0) &\mbox{or}&  (x\leq y<z \,\mbox{,} \,\, \hat{y}=1)
    \ea
    $$ 
    will be denoted by  $Plac_{\mathbb Z_2}(V)$, a super-counterpart of the plactic algebra.
These relations are a $\mathbb Z_2$-version of the Knuth relations\footnote{The super-Knuth relations obtained in the work \cite{superpl} are the same 
up the  sign depending on the $\mathbb Z_2$-grading. 
   }
of the plactic monoid on a signed alphabet $\{1, \ldots m, \bar{1}, \ldots, \bar{n}\}$ of the indices of the basis of the superspace $V=V_0\oplus V_1$. 
The classes of equivalent words modulo the super-Knuth relations are in one-to-one correspondence with the semistandard $(m,n)$-hook tableaux on one hand and
and to the states in the universal parastatistics algebra $\PS(V)$ on the other hand. 

The states in the parastatistics Fock space $\mathcal F(m|n,p)$ correspond to 
semistandard $(m,n)$-hook tableaux whose rows are $p$-restricted, that is,
with lengths not exceeding  $p$ boxes.
In the superalgebra $Plac_{\mathbb Z_2}(V)$ the $p$-restriction on the rows is imposed
by the condition
$$
 x_1 \ldots x_k y_{k+1} \ldots y_{p+1}=0 \qquad 
 x_1 \leq \ldots \leq x_k < y_{k+1}< \ldots <y_{p+1}
$$
where $\hat{x}_i =0$ and $\hat{y}_j=1$.
 

\section*{Acknowledgments}

It's a pleasure to thank Michel Dubois-Violette  for his constant interest  and encouragement.
He had first the idea to draw a parallel between the parastatistics algebra and the plactic monoid.
Todor Popov is indebted  to Boyka Aneva,  Oleg Ogievetsky, Tchavdar Palev, Neli Stoilova
and Joris Van der Jeugt for many enlightening discussions.
T.P. thanks for the hospitality of the Institut de Recherche Math\'ematique Avanc\'ee, Strasbourg and Centre de Physique Th\'eorique, Luminy  and acknowledges partial support from  the project  GIMP No.ANR-05-BLAN-0029-01 of the Agence Nationale pour la Recherche and 
the Bulgarian National
Foundation for Scientific Research (contract Ph-1406).
\appendix
    \section{Parafermionic Fock Space  
    Character}
    \label{O}
We are giving a sketch of the proof of the character identity  
  \beq
\label{pschar} 
\sum_{\eta \in F_p} (-1)^{\frac{1}{2}( |\eta|- (p-1)r)} s_{\eta}(x)=
{\prod_{i} {(1-x_{i})} 
\prod_{i<j}{(1-x_{i}x_{j}} )} 
\sum_{\lambda:\, l(\lambda')\leq p} s_{\lambda}(x) \ 
\eeq
which is equivalent to the one in eq.(\ref{identp}) when we change the summation
on the self-conjugated partitions $F_0$ by summation on the $p$-augmented partitions $F_p$.
In  Macdonald's book on symmetric functions the special case $p=0$ of the latter formula
 is given (see p.79 \cite{Macdonald})
$$ 
\sum_{\eta \in F_0} (-1)^{\frac{1}{2}(|\eta| +r)} s_{\eta}(x) =
{\prod_{i} {(1-x_{i})} 
\prod_{i<j}{(1-x_{i}x_{j}} )}$$
where the sum is over the selfconjugated partitions $\eta=\eta'$.
\noindent

We shall prove the character identity for every $p \in \mathbb N$
with the help of the Weyl identity for the Weyl groups $W$ of type $A_{n-1}$ and $B_n$
$$
\sum_{w\in W} \varepsilon(w) \, e^{w \rho} = \prod_{\alpha \in R^{+}}
 (e^{\frac{\alpha}{2}} - e^{-\frac{\alpha}{2}})
$$
where $\varepsilon(w)$ is the sign of the element
$w$ and $\rho$ is the Weyl vector $\rho=\frac{1}{2}\sum_{\alpha \in R^{+}} \alpha$.
The  exponentials of the weights  are formal.

Let $v_i$ be the standard basis of $\mathbb R^n$.
The root system of type $A_{n-1}$ and $B_n$ are 
$$ \Delta_0=\{ \, \mp v_i \pm v_j \, \}  \quad \subset \quad  \Delta= 
\{ \, \pm v_i \, , \, \pm v_i \pm v_j \, ,  \mp v_i \pm v_j\}  
 \qquad \quad 1\leq i<j \leq n$$ 
and the subsystems of the positive and negative roots are $\Delta^{\pm}_0 =\Delta_0 \cap \Delta^\pm$ and 
$$ \Delta^+=\{ \, v_i \, , \,   v_i \pm v_j \, \} \ , \qquad  \Delta^-=\{\,  -v_i \, , \, - v_i \pm v_j \, \}\, \ , \qquad \qquad 1\leq i<j \leq n  \ .$$
hence the  respective Weyl vectors differ by a constant vector $\theta=\frac{1}{2}\sum v_i$
$$\rho_0 = \sum_{i=1}^n (\frac{n}{2}+ \frac{1}{2}-i)v_i \ , \qquad \rho = \sum_{i=1}^n (n+ \frac{1}{2}-i)v_i \ , \qquad \rho = n\theta + \rho_0 \ .$$ 

The Weyl group $W_0$ of the root system $\Delta_0$ is the symmetric group $\S_n$ permuting the indexes of the vectors
$v_i$, while the Weyl group $W$ of the root system $\Delta$ is the semidirect product of $W_0$ with
the group $\mathbb Z_2^n$ acting by $v_i \mapsto \epsilon_i v_i$ where $\epsilon_i=\pm $.

The root system $\Delta$ has two commuting  involutions $c_0$ and $c$  
$$c_0 : v_i\mapsto v_{i'}:= v_{n+1-i} \ , \qquad \qquad  c : v_i \mapsto - v_i $$ 
uniquely determined by $c_0 (\Delta^+_0) = \Delta^-_0$ and $c (\Delta^+) = \Delta^-$.
These involutions are defined by the action of the element of the maximal length
in the Weyl group of $A_{n-1}$ and $B_n$, respectively.


The Weyl identity for $\Delta^+$ yields
$$ \sum_{w \in W} \varepsilon(w) e^{w\rho}  = 
e^\rho \prod_i (1 - e^{-v_i}) \prod_{i<j} (1- e^{-v_i -v_j})\prod_{i<j}  (1- e^{-v_i +v_j})  $$

The sum over the Weyl group $W$ can be split as
\beq
\sum_{\{ \epsilon_i \} \in \mathbb Z_2^n}\sum_{w_0 \in \S_n} (-1)^{\epsilon} 
\varepsilon(w_0) e^{w_0 \epsilon_i \rho}
= 
e^{n\theta}\sum_{\{\epsilon_i \}}  \sum_{ w_0 \in \S_n} (-1)^{\epsilon} 
 \varepsilon(w_0) 
e^{w_0 \rho_0 - w_0( \sum(1-\epsilon_i)\rho_i v_i)}
\eeq
where $\varepsilon(w_0)$ is the signature of the permutation $w_0$ and
$(-1)^{\epsilon} := \prod \epsilon_i 
=(-1)^{\sum\frac{1-\epsilon_i}{2}}$.

The action of the involution $c_0$  is equivalent to a resummation on $W_0$ 
and change the sum  by the signature of the element of maximal length in $W_0$, i.e., $(-1)^{\frac{n(n-1)}{2}}$. 

With the identification $x_i = e^{-v_i}$ and using the Weyl identity for $\Delta^+_0$
we get
\beq
\prod_i (1-x_i) \prod_{i<j}(1-x_i x_j) =
\sum_{\{ \epsilon_i \}}  (-1)^{\epsilon}
\frac{|x_j^{\mu_i} |}{|x_j^{\rho_{0i}} |} \quad \mbox{with} \quad  
\mu_i = \rho_{0i} + (1-\epsilon_{i'})\rho_{i'} \ .
\eeq

\begin{lemma}
\label{algo}
 In the $\S_n$-orbit of the weight $\mu =\mu(\epsilon_1, \ldots, \epsilon_n)=\sum \mu_i v_i$ uniquely determined
by the data of 
$ \{ \epsilon_1, \ldots, \epsilon_n \}$
 there exists one and only one
representative  
$$\lambda + \rho_0=\sigma(\mu) \qquad \mbox{such that}\qquad \lambda_1 \geq \lambda_2 \geq \ldots \geq \lambda_n \geq 0 \ ,$$
i.e., $\lambda$ is a Young diagram. 
The weight $\lambda$ is autoconjugated $\lambda=\lambda'$.
 The signature of the permutation $\sigma$ is
\beq
\varepsilon (\sigma)=(-1)^{\frac{1}{2} (\sum |\lambda| - r)} \ ,
\eeq
where $|\lambda|=\sum \lambda_i$ and $r$ is the number of diagonal boxes in $\lambda$.
\end{lemma}
{\it Proof:} By construction  the numbers $\mu_i$ 
 are all different
therefore $\mu_i$ can be always arranged by a permutation $\sigma$ into decreasing order, and then
 it will be a sum of a
partition $\lambda$ and
the  vector $\rho_0$ (with strictly decreasing components $\rho_{0i+1}-\rho_{0i} =1$). 

Given a  vector $\mu=\mu(\epsilon_1, \ldots, \epsilon_n)$, that is, a configuration
$ \{\epsilon_1, \ldots, \epsilon_n \}$ we choose  $\lambda$ to be such that  the projector $\frac{1-\epsilon_i}{2}$ projects to the hook  $(\beta_i|\alpha_i)$ in the Frobenius notation with $\alpha_i=\beta_i=n-i$.  In other words, $\lambda=(\alpha_1, \ldots, \alpha_r| \alpha_1, \ldots \alpha_r)$ and 
$r=\sum \frac{1-\epsilon_i}{2}$.
From the formula for the number of boxes of  $\lambda$
$$|\lambda|=\sum \mu_{\sigma(i)} -\sum  \rho_{0i}=\sum_i \rho_i(1-\epsilon_i)=\sum_i (2n -2i +1)\frac{1-\epsilon_i}{2} $$
it follows that our choice of $\lambda$ is a compatible one and therefore is the only possible one
since $\lambda$ such that $\lambda + \rho_0 = \sigma(\mu)$ is unique.

 For example the $2^2$ Young diagrams $\lambda$ appearing in the case $n=2$ read
 $$
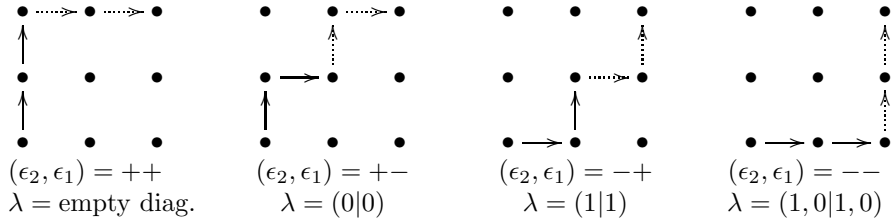
\ba{lccc}
\xymatrix @-0.8pc{ 
\bullet \ar@{.>}[r]& \bullet \ar@{.>}[r]& \bullet\\
\bullet \ar[u]& \bullet& \bullet\\
\bullet \ar[u]  & \bullet& \bullet
 }\quad &\quad
\xymatrix @-0.8pc{
\bullet & \bullet \ar@{.>}[r]& \bullet\\
\bullet \ar[r] & \bullet \ar@{.>}[u]& \bullet\\
\bullet \ar[u] & \bullet& \bullet
}\quad &\quad
\xymatrix @-0.8pc{
\bullet & \bullet& \bullet\\
\bullet & \bullet \ar@{.>}[r]& \bullet \ar@{.>}[u]\\
\bullet \ar[r] & \bullet \ar[u]& \bullet
}\quad &\quad
\xymatrix @-0.8pc{
\bullet & \bullet& \bullet\\
\bullet & \bullet& \bullet \ar@{.>}[u]\\
\bullet \ar[r] & \bullet \ar[r]& \bullet \ar@{.>}[u]
}\\[6pt]
(\epsilon_2,\epsilon_1)=++&(\epsilon_2,\epsilon_1)=+- & (\epsilon_2,\epsilon_1)=-+&
(\epsilon_2,\epsilon_1)=-- \\
\lambda = \mbox{empty diag.}&\lambda=(0|0)&\lambda=(1|1)&\lambda=(1,0|1,0)\ea
$$

The signature of $\sigma$ is the number of exchanges needed to bring the components
$\mu_i =\rho_{0i} + (1-\epsilon_{i'})\rho_{i'}$ into
 decreasing order,
 \beqa
 \varepsilon(\sigma)&=&\prod_{i<j} sgn(\mu_i - \mu_j) =
 \prod_{i<j} sgn ( (1-\epsilon_{i'})\rho_{i'} - (1-\epsilon_{j'})\rho_{j'} )\nn
&=&
 \prod_{i>j} sgn ( (1-\epsilon_{i})\rho_{i} - (1-\epsilon_{j})\rho_{j} ):=
 \prod_{i>j} s_{ij} \ .
 \nonumber
  \eeqa
Due to $\rho_i<\rho_j$ when $i>j$ the exchange sign  $s_{ij}$ depends only 
on the smaller index 
$$
s_{ij} =sgn((1-\epsilon_i)\rho_i - (1-\epsilon_j)\rho_j)
=(-1)^{\frac{1-\epsilon_j}{2}} \ , \qquad\qquad i>j
$$
hence the signature  of the permutation $\sigma$ depends only on $|\lambda|$ and $r$ 

$$
\varepsilon(\sigma) =\prod_{i>j} s_{ij}
 =(-1)^{\sum_j (n-j)\frac{1-\epsilon_j}{2}} = (-1)^{\frac{1}{2}(|\lambda|-r)} \ .
$$
$\Box$

Using the lemma and the determinantal  formula 
$s_{\lambda}(x){|x_j^{\rho_{0i}} |}={|x_j^{\rho_{0i} +\lambda_i} |}$ 
 we get
\beq
\prod_i (1-x_i) \prod_{i<j}(1-x_i x_j) =\sum_{\{\epsilon_i\} } (-1)^{\epsilon}
\frac{|x_j^{\mu_i} |}{|x_j^{\rho_{0i}} |}= \sum_{\lambda\in F_0} (-1)^{\frac{1}{2}(|\lambda|+r)}
s_{\lambda}(x)
\eeq
where the sum runs on the autoconjugated Young diagrams ($\mathfrak{gl}_n$-weights). 
This ends the proof of the character identity in the special case $p=0$. 

The sum over the Schur functions with no more than $p$ columns can be represented
as a quotient of determinants (see p.84 in the book of Macdonald \cite{Macdonald})
 $$
\sum_{\lambda:\, l(\lambda')\leq p} s_{\lambda}(e^{-v_1}, \ldots ,e^{-v_n})= e^{ -p \theta }D_{\rho  + p \theta} / D_{\rho}
$$
where  $D_{\rho}= \sum_{w\in W} \varepsilon(w) e^{w\rho}$.
The character  formula (\ref{pschar}) of the para-Fock module can be rewritten into the following equivalent form
\beq
\sum_{\eta \in F_p} (-1)^{\frac{1}{2}( |\eta|- (p-1)r)} s_{\eta}(x)
=   (-1)^{({}^n_2)} \, \, \frac{ e^{ -(n+p) \theta }D_{\rho  + p \theta} }{|x_j^{\rho_{0i}} |} \ .
\eeq

Proceeding as in the case $p=0$ we bring the RHS to
\beq
\sum_{ \{\epsilon_i\}}
(-1)^{\epsilon}
\frac{|x_j^{\nu_i} |}{|x_j^{\rho_{0i}} |} \quad \mbox{with} \quad  
\nu_i = \rho_{0i} + (1-\epsilon_{i'})\rho_{i'} + \frac{p}{2} \ .
\eeq

\begin{lemma} In the  $\S_n$-orbit of the weight $\nu =\nu(\epsilon_1, \ldots , \nu_n)=\sum \nu_i v_i$
there exists one and only one representative $\lambda$ such that 
$$
\lambda + \rho_0 = \sigma (\nu)\ , \qquad \qquad   \lambda_1 \geq \ldots \geq \lambda_n \geq 0 \ ,
$$
belongs to $F_p$, that is, $\lambda=(\alpha_1+p\, , \ldots , \alpha_r+p \, | \alpha_1, \ldots ,\alpha_r)$ with $r\leq n$.
 The signature of the permutation $\sigma$ is
$$
\varepsilon(\sigma)=(-1)^{\frac{1}{2}(|\lambda|-(p+1)r)} \ .
$$
The $2^n$ configurations of $\{ \epsilon_i \}$ are in bijection
with the $\mathfrak{gl}_n$-weights from $F_p$.  
\end{lemma}
{\it Sketch of the proof:} The proof goes in the same lines as the proof 
of lemma \ref{algo}.
 Given  a collection $\{ \epsilon_i \}$ we choose a diagram $\lambda$ such that 
the projector $\frac{1-\epsilon_i}{2}$ projects on the Frobenius hook $(\beta_i|\alpha_i)$
with  $\beta_i=n-i+p$ and $\alpha_i=n-i$. From the formula for 
the number of boxes in $\lambda\in F_p$ 
$$
|\lambda|=\sum (\nu_i-\rho_{0i})=\sum \rho_i(1-\epsilon_i)+ \frac{p}{2}=\sum (2n-2i+1 +p)\frac{1-\epsilon_i}{2} \ ,
$$
  implies 
$\lambda + \rho_0 = \sigma (\nu)$ (such a $\lambda$ is unique).
Every configuration $\{ \epsilon_i \}$ gives a different Young diagram $\lambda\in F_p$
thus the total number is $2^n$.
The signature $\varepsilon(\sigma)$ is 
$$
\varepsilon(\sigma)= \prod_{i<j} sgn (\nu_i - \nu_j) = (-1)^{\sum (n-j)\frac{1-\epsilon_j}{2}}= (-1)^{\frac{1}{2}(|\lambda|-(p+1)r)} \ .
$$ 
$\Box$

Finally with the help of the lemma we conclude that  
$$
\sum_{\{\epsilon_i\}} (-1)^{\epsilon}
\frac{|x_j^{\nu_i} |}{|x_j^{\rho_{0i}} |} 
= \sum_{\lambda \in  F_p} (-1)^r (-1)^{\frac{1}{2}(|\lambda|-(p+1)r)} 
\frac{|x_j^{\rho_{0i} + \lambda} |}{|x_j^{\rho_{0i}} |}=
\sum_{\lambda\in F_p} (-1)^{\frac{1}{2}(|\lambda|-(p-1)r)} s_{\lambda}(x)
$$
so the character identity (\ref{pschar}) of the parastatistics Fock space holds true.

\end{document}